\documentclass[sigconf,nonacm]{acmart}
\usepackage{popets}
\usepackage{microtype}
\usepackage[T1]{fontenc}
\usepackage{graphicx}

\usepackage{hyperref}
\usepackage{cleveref}
\usepackage{color}

\usepackage{tikz}

\definecolor{minTop}{HTML}{A9C4EB}       \definecolor{minBot}{HTML}{9999FF}
\definecolor{strTop}{HTML}{67AB9F}       \definecolor{strBot}{HTML}{CCFF99}
\definecolor{selTop}{HTML}{EA6B66}       \definecolor{selBot}{HTML}{FF9999}
\definecolor{destroyFill}{HTML}{FFCE9F}  \definecolor{destroyLine}{HTML}{C73500}
\definecolor{excludeFill}{HTML}{FFE599}  \definecolor{excludeLine}{HTML}{B09500}
\definecolor{leafGreen}{HTML}{E6FFCC}    \definecolor{leafRed}{HTML}{FFCCCC}
\definecolor{txtGreen}{HTML}{336600}     \definecolor{txtRed}{HTML}{6C0507}
\definecolor{txtNavy}{HTML}{000066}      \definecolor{txtMin}{HTML}{004C99}

\newcommand{\du}[1]{\dimexpr#1pt*3/4\relax}

\begin{document}

\title[On the Selection of Mitigations for GenAI Privacy Threats]{The Right Tool for the Job: On the Selection of Mitigations for GenAI Privacy Threats}

\author{Jonah Bellemans$^*$}
\orcid{0009-0006-4336-6370} \affiliation{\institution{DistriNet, KU Leuven}
  \city{Heverlee}
  \country{Belgium}}
\email{jonah.bellemans@kuleuven.be}

\author{Qianying Liao$^*$}
\orcid{0000-0002-5333-3103} \affiliation{\institution{DistriNet, KU Leuven}
  \city{Heverlee}
  \country{Belgium}}
\email{qianying.liao@kuleuven.be}

\author{Laurens Sion}
\orcid{0000-0002-8126-4491} \affiliation{\institution{DistriNet, KU Leuven}
  \city{Heverlee}
  \country{Belgium}
}
\email{laurens.sion@kuleuven.be}

\author{Lieven Desmet}
\orcid{0000-0001-5155-7472} \affiliation{\institution{DistriNet, KU Leuven}
  \city{Heverlee}
  \country{Belgium}}
\email{lieven.desmet@kuleuven.be}

\author{Wouter Joosen}
\orcid{0000-0002-7710-5092} \affiliation{\institution{DistriNet, KU Leuven}
  \city{Heverlee}
  \country{Belgium}}
\email{wouter.joosen@kuleuven.be}

\renewcommand{\shortauthors}{Bellemans et al.}

\begin{abstract}
 Generative Artificial Intelligence (GenAI) has rapidly evolved from an experimental technology into a foundational component of modern software systems. 
However, as its adoption grows, protecting sensitive personal data becomes increasingly challenging. 
Specifically, GenAI systems not only amplify traditional privacy threats but also introduce new inference-based risks, such as constructing detailed user profiles from seemingly harmless inputs.
In response, privacy threat modeling frameworks are beginning to capture GenAI-specific privacy threats with finer granularity. 
At the same time, a growing number of mitigation techniques have been proposed to address these threats. 
However, although knowledge of both threats and mitigations continues to mature, the problem- and solution-space have developed largely independently.

This position paper argues that the primary challenge in GenAI privacy engineering is not the lack of knowledge about privacy threats or mitigation techniques, but the missing bridge between them. We decompose this gap into three sub-problems: 
(i)~lack of fine-grained threat-to-mitigation mapping for GenAI systems, 
(ii)~inapplicable solution-space assumptions in the GenAI
context, and 
(iii)~prioritization difficulty under GenAI constraints. 
We derive four recommendations for future mitigation-selection approaches, and outline a suggested approach that extends established threat-to-mitigation mapping methods to GenAI-specific threat characteristics. 
We propose a research agenda toward more systematic privacy mitigation selection for GenAI-based systems.
\end{abstract}


\maketitle
\def\thefootnote{*}\footnotetext{These authors contributed equally to this work.}\def\thefootnote{\arabic{footnote}}
\section{Introduction}\label{sec:introduction}
Generative Artificial Intelligence (GenAI) systems are increasingly embedded in contemporary software systems, reshaping how data is processed, inferred, and exposed across a wide range of application domains.
Compared with earlier generations of AI systems, which were often developed for specialized tasks and deployed in tightly controlled environments, GenAI technologies significantly lower the barrier to adoption and integration.
Through widely accessible APIs, software development kits, and foundation model services, developers can rapidly incorporate GenAI capabilities into existing applications with minimal expertise.

While the widespread accessibility of GenAI technologies has accelerated their adoption across the software ecosystem, it has also propagated privacy and identity-related risks into a broader range of applications and deployment contexts. These risks span the entire GenAI lifecycle: models memorize and regurgitate training data, prompts and retrieval pipelines leak personal information, and models infer sensitive attributes from seemingly innocuous inputs~\cite{carlini2021,bodea2026,staab2024beyond}. 

A recent study found that ``GPT-4 and ChatGPT reveal private information in contexts that humans would not, 39\% and 57\% of the time, respectively''~\cite{mireshghallah2024}. Furthermore, these risks have already materialized in everyday applications and enterprise systems.
For instance, studies show that employees frequently submit sensitive corporate information to GenAI systems~\cite{muncaster2024}, with nearly 10\% of enterprise GenAI prompts containing confidential data such as source code, financial records, or customer information~\cite{schuman2025}.
In addition, GenAI systems have generated fabricated, privacy-invasive outputs involving real individuals, including a widely reported case in which ChatGPT falsely accused a law professor of sexual harassment~\cite{verma2023}.
Such findings raise significant concerns regarding the privacy implications of GenAI systems, particularly in enterprise environments processing identity-linked and organizational data.

The growing body of research on GenAI privacy describing the \emph{problem-space} led to increasingly
fine-grained methods for eliciting, characterizing, and cataloging
threat scenarios~\cite{bodea2026,das2025,ma2025,liao2026}.
Among these efforts, LINDDUN4GenAI~\cite{liao2026} extends the LINDDUN privacy
threat modeling framework with GenAI-specific threat characteristics and
Common Attacker Models (CAMs).

To address these risks, the \emph{solution-space} has evolved just as rapidly. 
Established resources, including privacy design strategies and tactics~\cite{hoepman2014,hoepman2022,colesky2016},
privacy patterns~\cite{privacypatterns}, and privacy-enhancing technologies (PETs)~\cite{heurix2015}, have been complemented by a growing body of GenAI-specific mitigation techniques~\cite{shanmugarasa2025,bodea2026}.

Although both the problem-space and solution-space continue to mature, a more fundamental challenge is that they have largely evolved independently from one another.
Consequently, the guidance linking identified threats to appropriate mitigations remains predominantly limited to high-level LINDDUN threat types, rather than leveraging the fine-grained threat characteristics captured during problem-space analysis.
As demonstrated by Al-Momani et al.~\cite{almomani2022} for traditional LINDDUN, much of the threat detail obtained during problem-space analysis is effectively \emph{lost in translation} when transitioning to the solution-space.
This disconnect is even more pronounced in the
GenAI domain, where many new GenAI-specific threat characteristics lack
targeted mitigation guidance altogether. 
Furthermore, even when multiple candidate
mitigations are applicable, there is no principled basis for prioritizing
among them under GenAI-specific privacy requirements and constraints.
Therefore, software engineers lack actionable and systematic support for
bridging GenAI-specific privacy threats to appropriate mitigations and for
evaluating alternative mitigations in a principled manner.

\emph{We argue that GenAI privacy engineering lacks neither threat knowledge
nor mitigation techniques, but rather the bridge between them: as long as both
continue to grow as separate bodies of knowledge and matching the right
mitigation to an identified threat remains unsupported by concrete, fine-grained selection methodologies, ever-finer threat modeling will not yield more private systems.}

We concretize this gap by discussing the current state of the art and where it falls short. We explicitly present three concrete sub-problems and illustrate these based on a running example of an HR chatbot system.

Addressing this gap requires GenAI-specific mitigation-selection approaches that do not yet
exist. Rather than prescribing one, we formulate four recommendations
that such approaches should satisfy, and suggest one potential approach:
extending the more general methodology of Al-Momani et al.~\cite{almomani2022} to the GenAI-specific threat characteristics
introduced by LINDDUN4GenAI~\cite{liao2026}.
Where relevant, such guidance should also be informed by legal obligations
under frameworks such as the EU AI Act~\cite{EU2024ai}, so that selected
mitigations can be situated within applicable regulatory expectations and
data subject rights.
In doing so, the paper outlines a research agenda for bridging the problem- and
solution-space of GenAI privacy engineering.

The remainder of this paper is structured as follows.
\Cref{sec:background} provides background and discusses related work on privacy threat modeling, GenAI system paradigms,
LINDDUN4GenAI and its threat characteristics, the layers of the
solution-space: privacy design strategies and tactics, privacy patterns,
and PETs, and finally existing efforts to bridge the gap between the problem- and the solution-space.
\Cref{sec:problem} elaborates on the gap statement and explains the three core sub-problems through a running example.
\Cref{sec:approach} formulates recommendations for future
mitigation-selection approaches and sketches a candidate direction extending
the approach of Al-Momani et al.~\cite{almomani2022} to the GenAI context.
\Cref{sec:conclusion} concludes with a call to action for the community.
 \section{Background and Related Work}\label{sec:background}
This section introduces the concepts which the remainder of this paper is built on: the practice of privacy threat modeling and the LINDDUN framework (\cref{sec:bg-linddun}), its GenAI-specific extension LINDDUN4GenAI and the threat characteristics central to our argument (\cref{sec:bg-genai}), the layered solution-space of strategies, tactics, patterns, and PETs from which mitigations are drawn (\cref{sec:bg-solutionspace}), and finally the available guidance to bridge the gap between threats and their suitable mitigations (\cref{sec:bg-threatmitigationmapping}).

\subsection{Privacy threat modeling}\label{sec:bg-linddun}
Threat modeling is a systematic approach to identifying what can go wrong in a system before it is built~\cite{shostack2014}. Originating in security engineering, approaches such as STRIDE~\cite{kohnfelder1999} enumerate threat categories (or `\emph{types}') over a model of the system under design. Over time, a broad family of variants of such methods has emerged~\cite{xiong2019}, including frameworks for dedicated privacy threat modeling.

\begin{figure}[!ht]
    \centering
\includegraphics[width=0.95\linewidth]{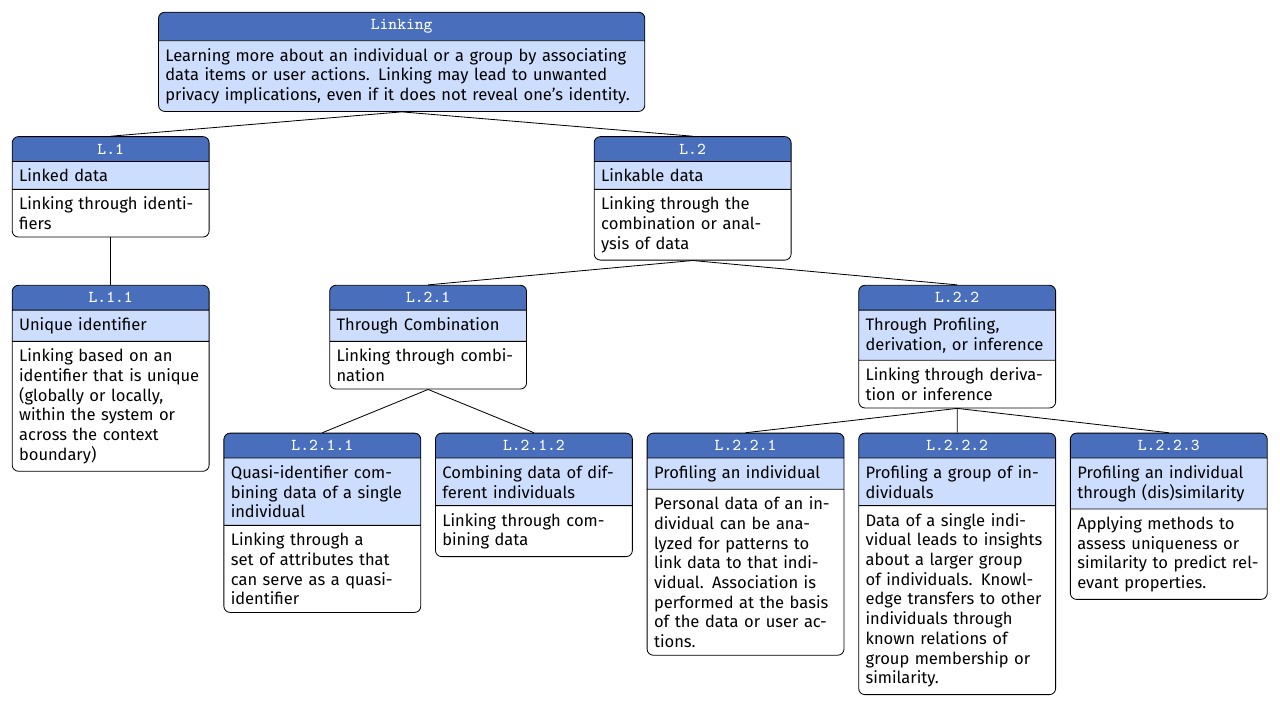}
\caption{Linking threat tree}\label{fig:linking}
\end{figure}

Among these frameworks, LINDDUN~\cite{deng2011,LINDDUNwebsite} is one of the most widely used. Starting from a model of the system under design, typically a Data Flow Diagram (DFD), it guides analysts through the elicitation of privacy threats in the seven threat types of its mnemonic: Linking, Identifying, Non-repudiation, Detecting, Data Disclosure, Unawareness \& Unintervenability, and Non-compliance. The first five are commonly referred to as \emph{hard} privacy threats, concerning properties such as anonymity, unlinkability, and confidentiality; the latter two as \emph{soft} privacy threats, concerning transparency, intervenability, and compliance~\cite{deng2011}. For each privacy threat type, LINDDUN provides a threat tree that refines the threat type into \emph{threat characteristics}: the lower the node, the more specific the description of how, and due to which cause, the threat arises. An example of such a threat tree can be seen in \cref{fig:linking}~\cite{LINDDUNwebsite}. This information is stored in an actively maintained and reusable threat knowledge base~\cite{sion2025}.

Complementary frameworks approach privacy threats from other angles. MITRE's PANOPTIC~\cite{shapiro2023} offers an alternative privacy threat model rooted in a taxonomy of privacy threat activities; PRIAM~\cite{de2016} complements threat elicitation with privacy \emph{risk} analysis, quantifying the severity of identified risks; and PLOT4AI~\cite{plot4ai} provides a practical library of threat cards for AI systems. These efforts differ in elicitation style and scope, but share LINDDUN's focus on the problem-space.

\subsection{GenAI-based systems and LINDDUN4GenAI}\label{sec:bg-genai}
Users and organizations interact with GenAI in four main deployment paradigms: (i)~direct use of a pre-trained foundation model, (ii)~interaction with a fine-tuned model adapted to a domain, (iii)~use of an application built around a model through system prompts, and (iv)~agentic systems in which the model is embedded in a broader software ecosystem and invokes external tools and data sources~\cite{liao2026}. The paradigms differ in who controls the training data, the model weights, and the prompts, a distinction that becomes central to mitigation feasibility later in this paper.

LINDDUN4GenAI~\cite{liao2026} extends the LINDDUN threat knowledge base to these systems. From a synthesis of the academic literature and two industrial case studies (a GenAI-based chatbot and an agentic assistant), it identifies six Common Attacker Models (CAMs): recurring types of leakage defined by the targeted information and the acting party. CAM1 and CAM2 concern user-to-system and system-to-user information leakage, while CAM3 covers leakage of information from the pre-training to the fine-tuning party. CAM4 and CAM5 concern leakage between the system and agents (in either direction). Finally, in CAM6 (\emph{residual privacy leakage}), a party with access to the model's intermediate computations, such as embeddings, activations, or gradients, can invert them to recover personal data that was never disclosed in any output~\cite{liao2026}.

\begin{figure}[ht]
    \centering
    \includegraphics[width=0.95\linewidth]{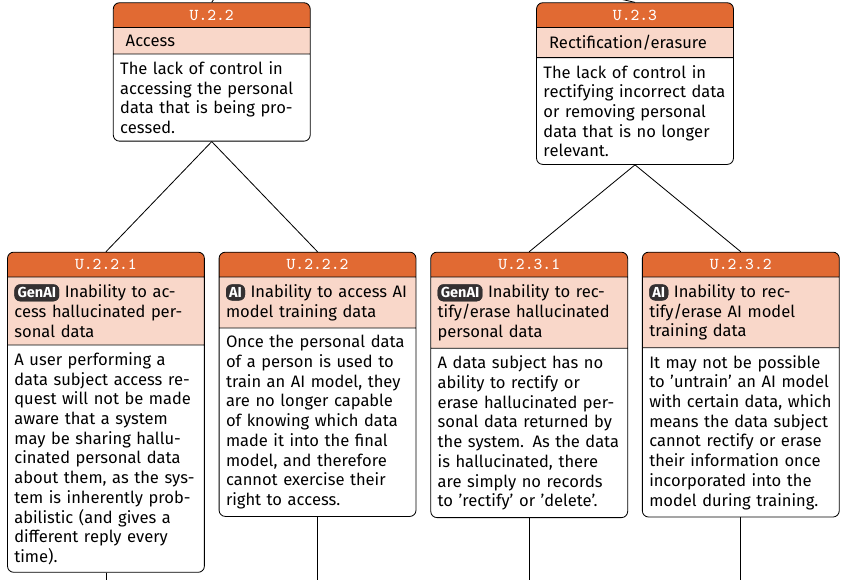}
    \caption{GenAI-specific threat tree nodes}\label{fig:unawareness-genai}
\end{figure}

Based on these CAMs, LINDDUN4GenAI adds nine new GenAI-specific threat characteristics to the threat trees. Several of these are central to this paper. 
Among the hard privacy characteristics, \emph{DD.1.3 Data type structure} captures disclosure through the structure of data representations themselves; its child \emph{DD.1.3.2 Data derivations} covers the model-internal derivations exploited in CAM6. \emph{DD.3.5 Fabrication} captures the disclosure of incorrect, hallucinated personal data that is presented as if it were accurate. 

The existing soft privacy characteristics \emph{U.2.2 Access} and \emph{U.2.3 Rectification/erasure} are refined with AI-specific children (as illustrated in \cref{fig:unawareness-genai}~\cite{liao2026}) reflecting that hallucinated personal data leave no records that could be accessed, rectified, or erased, and that data absorbed into model weights cannot, in general, be selectively untrained. Finally, \emph{U.3 Interference with personal decision making} introduces an entirely new class of threat: the system interfering with the autonomy of the data subject, for example through manipulation. Two further characteristics concern non-compliance with the EU AI Act (Nc.1.3) and with AI standards (Nc.4.2).

\subsection{Privacy mitigations: strategies, tactics, patterns, and PETs}\label{sec:bg-solutionspace}
\begin{figure*}
  \centering
\resizebox{0.7\linewidth}{!}{\begin{tikzpicture}[x=0.75pt,y=-0.75pt,
bx/.style n args={2}{anchor=north west,minimum width=\du{#1},minimum height=\du{#2},
                       align=center,inner sep=0pt,outer sep=0pt,line width=0.75pt},
tA/.style  ={font=\fontsize{9}{10.8}\selectfont,   text height=7.13pt, text depth=2.25pt},
  tAii/.style={font=\fontsize{9}{10.8}\selectfont,   text height=19.94pt,text depth=2.25pt},
  tB/.style  ={font=\fontsize{10.5}{12.6}\selectfont,text height=10.185pt,text depth=2.625pt},
  tBb/.style ={font=\fontsize{10.5}{12.6}\bfseries\selectfont,text height=10.185pt,text depth=2.625pt},
  tC/.style  ={font=\fontsize{13.5}{16.2}\bfseries\selectfont,text height=13.095pt,text depth=3.375pt},
  leader/.style={line width=0.75pt,dash pattern=on 2.25pt off 2.25pt}]

\draw[leader] (772.46,356) -- (880,356);

\path[top color=minTop,bottom color=minBot] (380,230) -- (724,230) -- (807,410) -- (297,410) -- cycle;
\path[top color=selTop,bottom color=selBot] (568,305) -- (718,305) -- (768,400) -- (518,400) -- cycle;
\path[top color=strTop,bottom color=strBot] (386,305) -- (536,305) -- (586,400) -- (336,400) -- cycle;

\node[bx={344}{39.7},fill=minTop,text=txtMin,tC] at (380,200.3) {Minimize};

\node[bx={150}{30},fill=destroyFill,draw=destroyLine,tB] at (386,240) {Destroy};
\node[bx={150}{30},fill=excludeFill,draw=excludeLine,tB] at (568,240) {Exclude};
\node[bx={150}{30},fill=strTop,text=white,tBb]           at (386,275) {Strip};
\node[bx={150}{30},fill=selTop,text=white,tBb]           at (568,275) {Select};

\node[bx={123}{41},fill=leafGreen,draw=black,text=txtGreen,tAii] at (399.5,310) {Strip Invisible\\Metadata};
\node[bx={120}{40},fill=leafRed,draw=black,text=txtRed,tA]       at (586,311)   {Selective Disclosure};

\node[bx={75}{35},fill=white,draw=black,text=txtGreen,tAii] at (381,357.5) {PII detection\\and redaction};
\node[bx={75}{35},fill=white,draw=black,text=txtGreen,tAii] at (464,357.5) {Metadata\\sanitization};
\node[bx={75}{35},fill=white,draw=black,text=txtRed,tAii]   at (570,357.5) {Query\\rewriting};
\node[bx={75}{35},fill=white,draw=black,text=txtRed,tAii]   at (652,357.5) {Context\\filtering};

\draw[leader] (714,200.8)  -- (760,200.8);
\draw[leader] (718,240)    -- (800,240);
\draw[leader] (755,310)    -- (843.42,310);

\node[text=txtNavy,tB,inner sep=0pt] at (812.5,211)   {Privacy Strategies};
\node[text=txtNavy,tB,inner sep=0pt] at (823,249.3)   {Privacy Tactics};
\node[text=txtNavy,tB,inner sep=0pt] at (819,322)     {Privacy Patterns};
\node[text=txtNavy,tB,inner sep=0pt] at (852,368.5)   {PETs};
\end{tikzpicture}}
\caption{Hierarchy of privacy solution-space knowledge illustrated using the \emph{Minimize} privacy strategy as an example.}\label{fig:privacy-hier}
\end{figure*}
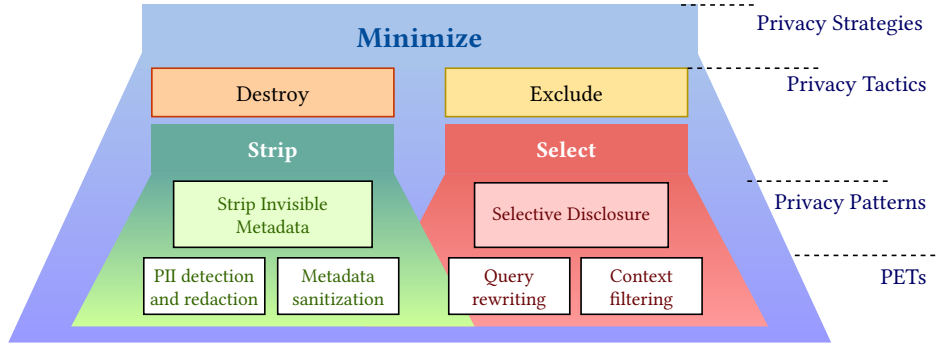

The solution-space knowledge available to privacy engineers is organized into four layers of decreasing abstraction, as shown in~\cref{fig:privacy-hier}.

At the top, \emph{privacy design strategies} express architectural goals: Hoepman~\cite{hoepman2014,hoepman2022} distinguishes four data-oriented strategies (minimize, separate, abstract\footnote{Hoepman initially named this strategy `aggregate'~\cite{hoepman2014}, but it was later renamed by Colesky et al.~\cite{colesky2016}. The new naming is also reflected in Hoepman's little blue book~\cite{hoepman2022}.}, hide) and four process-oriented ones (inform, control, enforce, demonstrate). 

For the second layer, Colesky et al.~\cite{colesky2016} refine each strategy into \emph{tactics}, a refinement later consolidated in Hoepman's \emph{little blue book}~\cite{hoepman2022}. In the example of \cref{fig:privacy-hier}, the \emph{Minimize} strategy is refined into the tactics \emph{Destroy}, \emph{Exclude}, \emph{Strip}, and \emph{Select}.

At the third layer, \emph{privacy patterns} describe reusable solutions to recurring design problems and are compiled in a community catalog~\cite{privacypatterns}, classified by the strategies and tactics they realize.

At the bottom, \emph{privacy-enhancing technologies} (PETs) are concrete techniques implementing (parts of) patterns, ranging from attribute removal to differential privacy and homomorphic encryption, and have been organized in dedicated taxonomies~\cite{heurix2015}. Moving down the layers, guidance becomes more concrete but also more context-dependent: which PET is appropriate depends on properties of both the system and the threat, which is where the selection problem highlighted by this paper arises.

As illustrated in \cref{fig:privacy-hier}, the \emph{Strip} tactic is exemplified by the \emph{Strip Invisible Metadata} pattern, which may be implemented using PETs such as PII (personally identifiable information) detection and redaction or metadata sanitization. Likewise, the \emph{Select} tactic is illustrated by the \emph{Selective Disclosure} pattern, which may be realized using PETs such as query rewriting and context filtering (note: the listed PETs here are merely illustrative, they do not form a complete mapping). The hierarchy demonstrates the increasing level of implementation specificity from strategies to PETs and motivates the solution-space organization.

\subsection{Mitigation selection guidance}\label{sec:bg-threatmitigationmapping}
A first line of work supports the selection of mitigations without starting from elicited privacy threats.
Kunz and Binder~\cite{kunz2022} observe that existing PET systematizations are outdated or focus on comparison criteria rather than on guidance for practical selection, and
propose a classification that adds functional context, technology maturity, and impact on non-functional requirements to existing criteria, such as the privacy protection goal.
Pape et al.~\cite{pape2025} anchor mitigation selection in data protection law instead, mapping trust models and PET types onto GDPR principles. Baldassarre et al.~\cite{baldassarre2022} formalize the relations among privacy-by-design principles, design strategies, patterns, and vulnerabilities in a privacy knowledge base with accompanying tool support. The entry point of these approaches is the functional context of the system, a legal principle, or a vulnerability, rather than a privacy threat identified during analysis. The fine-grained, specific threat information obtained during threat elicitation therefore plays no role in the selection of suitable mitigations with these techniques.

A selection methodology that does take the elicited threat information into account was proposed by the authors of LINDDUN itself in its earlier stages~\cite{deng2011}. Deng et al.\ accompany the threat trees with a taxonomy of mitigation strategies and a classification of privacy-enhancing solutions, linking each threat type to a mitigation strategy and, through it, to candidate PETs~\cite{deng2011}. However, the mapping operates at the level of threat types, not the more fine-grained threat characteristics. For example, it indicates what to do about identifying threats in general, but not what to do about specific causes of identifying threats elicited within the system. Furthermore, the LINDDUN threat knowledge has since been reworked multiple times~\cite{sion2025}, as well as extended to GenAI~\cite{liao2026}, without a corresponding update on the mitigation side.

On the GenAI side, the closest counterpart to such a mapping is a survey by Wang et al.~\cite{wang2025}. It divides the LLM lifecycle into four scenarios, namely pre-training, fine-tuning, deployment, and LLM-based agents, which parallel the deployment paradigms mentioned by LINDDUN4GenAI~\cite{liao2026}, pairing the risks of each scenario with countermeasures. Their comparison tables record, for each countermeasure, the risk it aims to mitigate, the access the defender would require to the model and the training data, the models and tasks to which it applies, and its reported limitations. However, the countermeasures are once again mapped to higher-level risks, rather than specific threat characteristics attached to a model of the system under design.
Other SoK papers follow the same pattern: they pair risks with mitigations for a single paradigm or a single class of techniques, and assess maturity rather than fit to an identified threat~\cite{bodea2026,shanmugarasa2025}.

To our knowledge, the only works that bridge fine-grained classical LINDDUN threats to concrete mitigations are those of Al-Momani et al.\footnote{While these works all share the same first author, we note that the co-authors involved only partially overlap and may differ per publication.}~\cite{almomani2022,almomani2021,almomani2024,almomani2026}.
First, they analyze 70 patterns of a widely used privacy pattern catalog along five property dimensions that govern selection (applicability scope, privacy objective, impacted qualities, data focus, and hotspot), and find that these properties are rarely made explicit in the catalog itself~\cite{almomani2021}. Subsequently, they show that the fine-grained knowledge gained during threat analysis is \emph{lost in translation} from the problem-space to the solution-space, and propose a remedy: 36 \emph{key nodes} identified in the LINDDUN threat trees, grouped into ten mitigation goals, with solution flowcharts for the four goals that admit many candidate countermeasures~\cite{almomani2022}. In later experiments, they evaluate and build further upon this work in practice. Eliciting threats for a robotaxi service with LINDDUN and matching them against the pattern catalog, they report inconsistencies within the patterns, a lack of guidance on applying these patterns, as well as threats for which no suitable pattern exists~\cite{almomani2024}. Applying three PET-selection approaches (including their own) to the same setting, they conclude that none adequately supports selection in a realistic scenario, and derive requirements for future selection methodologies~\cite{almomani2026}. \section{Problem Statement}\label{sec:problem}

We begin by analyzing the gap between the increasingly fine-grained GenAI threat knowledge available in the problem-space, such as LINDDUN4GenAI~\cite{liao2026}, and the mitigation guidance available in the solution space (\Cref{subsec:gap}). 
We then introduce a running example of an HR chatbot to illustrate the limitations of existing threat-to-mitigation guidance in a realistic setting (\Cref{subsubsec:example}). 
Finally, we decompose this gap into three concrete sub-problems in~\cref{subsec:subproblems}: 
(i)~mapping GenAI threats to mitigations at an appropriate level of granularity; 
(ii)~determining whether existing mitigations remain applicable in the GenAI context; and 
(iii)~prioritizing among multiple applicable mitigations.

\subsection{Research Gap}
\label{subsec:gap}
The gap this paper addresses is best characterized by highlighting where the current state of the art falls short.

On the problem-space side, 
LINDDUN4GenAI~\cite{liao2026} extends the LINDDUN threat trees with GenAI-specific threat characteristics and Common Attacker Models, 
but deliberately stops at threat elicitation: 
it prescribes neither which mitigation addresses a given threat, nor how to choose among alternatives. As mentioned in \cref{sec:bg-threatmitigationmapping}, the existing threat-to-mitigation mapping of LINDDUN~\cite{deng2011} is insufficiently fine-grained, has not been updated to reflect the latest knowledge base contents and structure, and predates the GenAI threat characteristics.
Adjacent security-oriented resources do not fill this gap either. 
The OWASP Top~10 for LLM Applications~\cite{OWASPTTLLM} and MITRE ATLAS~\cite{ATLAS} index mitigations by vulnerability and adversarial technique, not by privacy threat.

On the solution-space side, 
we examine the line of work of Al-Momani et al.~\cite{almomani2021,almomani2022,almomani2024,almomani2026} in more detail, since it is the only one connecting fine-grained LINDDUN threat knowledge to concrete mitigations, and since it reports on applying that connection to a real-world case.

Even for classical LINDDUN, their results leave the solution-space unevenly covered from two different perspectives. 
On the side of the pattern catalogs, 
they show that the available patterns predominantly serve soft privacy goals such as transparency and intervenability, while far fewer address hard privacy goals such as anonymity and unlinkability~\cite{almomani2021}. In later work, they further quantify this imbalance, reporting 59 process-oriented patterns versus 19 data-oriented patterns\footnote{In this work, Al-Momani et al.\ report more than 70 patterns, indicating that the pattern catalog has grown since their ``Land of the Lost'' paper in 2021~\cite{almomani2021}.}~\cite{almomani2024}.
On the guidance side, their decision support~\cite{almomani2022} covers only the five hard privacy threat types and explicitly excludes the soft ones, on the grounds that selection support for soft privacy countermeasures has been treated elsewhere in the form of pattern systems. \textbf{Each group therefore lacks what the other has: structured selection guidance for hard privacy threats, but few patterns to select from; many patterns for soft privacy threats, but guidance organized by design strategy rather than by elicited threat.} Their robotaxi case study~\cite{almomani2024} indicates that this leaves engineers without a usable route from a threat to a pattern. They report inconsistent levels of abstraction across patterns and a lack of guidance for finding the right pattern for a given type of threat. In addition, they had to propose new patterns for threats the catalog does not address at all, particularly for data deletion and replacing sensitive data~\cite{almomani2024}.

The two most recent papers also evaluate their methodology in practice, and report that it still has shortcomings which need to be addressed. Matching each elicited threat of the robotaxi design against the pattern catalog, \textbf{they find that mitigating a single threat frequently requires a combination of patterns for which no composition guidance exists, and that even once a suitable pattern has been identified, selecting a concrete PET to realize it remains difficult~\cite{almomani2024}.} The most recent study applies three published PET-selection approaches~\cite{almomani2022,kunz2020,kunz2022}, complemented by a pragmatic approach based on privacy design strategies~\cite{hoepman2014}, to the same use case, and concludes that none of them adequately supports PET selection in a complex real-world scenario~\cite{almomani2026}. They make three observations relevant for the argument of this paper. First, the approaches yield a set of applicable PETs but offer little support for the final choice among them. Second, each approach treats threats in isolation and selects at least one PET per threat, whereas both threats and PETs can influence each other in practice (e.g., the introduction of a PET might also introduce new threats to the system). Third, the solution flowcharts of Al-Momani et al.~\cite{almomani2022} prove sparsely populated: the chart for protecting identity lists attribute-based credentials as the only countermeasure, and the chart for protecting attributes refers to encryption in general without naming a specific technology.
Finally, the line of work of Al-Momani et al.\ predates the GenAI-specific threat knowledge base of LINDDUN4GenAI.\ The pattern properties of Al-Momani et al.~\cite{almomani2021} are defined for traditional software architectures, and their robotaxi studies concern a cyber-physical system in which machine learning is a component of the use case rather than the object of the privacy analysis~\cite{almomani2024,almomani2026}.
Their key nodes and flowcharts~\cite{almomani2022} rest on an earlier generation of the LINDDUN threat trees. The structure and node identifiers of that generation differ from the restructured knowledge base on which LINDDUN4GenAI builds. This issue was explicitly mentioned in their follow-up case study, in which they mention that their approach had to be modified because it had been designed for this earlier version of LINDDUN, rendering the use of the key nodes unfeasible~\cite{almomani2026}.
\textbf{None of these results covers any of the GenAI-specific threat characteristics. They are thus not tailored to the problem- and solution-space that GenAI introduces.}

\subsection{Running Example}
\label{subsubsec:example}
To make these limitations concrete, we use an LLM-based HR chatbot as a running example (depicted schematically in \cref{fig:running-example}).

\definecolor{hrcBlue}{HTML}{1F6D91}     \definecolor{hrcMagenta}{HTML}{A02B93}  \definecolor{hrcTeal}{HTML}{127A68}     \definecolor{hrcGreen}{HTML}{5F7C1F}    \definecolor{hrcArrow}{HTML}{BFC2C2}
\definecolor{hrcArrowEdge}{HTML}{6E7273}
\definecolor{hrcText}{HTML}{1A1A1A}

\tikzset{
  hrc icon user/.pic={
  \path[fill=hrcBlue]
    (0.4688,-0.2656) .. controls (0.4672,-0.2266) and (0.4484,-0.1875) .. (0.4156,-0.1609) .. controls (0.3734,-0.1250) and (0.3156,-0.0922) .. (0.2375,-0.0609) .. controls (0.2812,-0.0500) and (0.3188,-0.0266) .. (0.3203,-0.0250) .. controls (0.3250,-0.0219) and (0.3266,-0.0188) .. (0.3281,-0.0141) .. controls (0.3281,-0.0094) and (0.3266,-0.0047) .. (0.3234,-0.0016) .. controls (0.3234,-0.0016) and (0.2750,0.0453) .. (0.2703,0.0953) .. controls (0.2688,0.1156) and (0.2766,0.1578) .. (0.2859,0.2031) .. controls (0.2969,0.2609) and (0.3094,0.3266) .. (0.3094,0.3797) .. controls (0.3094,0.4562) and (0.2891,0.4984) .. (0.2391,0.5281) .. controls (0.2172,0.5406) and (0.1797,0.5359) .. (0.1609,0.5328) .. controls (0.1469,0.5516) and (0.1109,0.5891) .. (0.0344,0.6141) .. controls (-0.0484,0.6359) and (-0.1344,0.6250) .. (-0.2094,0.5828) .. controls (-0.2766,0.5391) and (-0.3047,0.4703) .. (-0.3047,0.3469) .. controls (-0.3047,0.2859) and (-0.3016,0.2328) .. (-0.2984,0.1859) .. controls (-0.2953,0.1328) and (-0.2922,0.0906) .. (-0.2969,0.0625) .. controls (-0.3000,0.0344) and (-0.3156,0.0094) .. (-0.3312,-0.0109) .. controls (-0.3344,-0.0141) and (-0.3344,-0.0188) .. (-0.3344,-0.0234) .. controls (-0.3344,-0.0281) and (-0.3312,-0.0312) .. (-0.3266,-0.0344) .. controls (-0.3234,-0.0359) and (-0.2891,-0.0547) .. (-0.2453,-0.0641) .. controls (-0.3219,-0.0938) and (-0.3781,-0.1266) .. (-0.4188,-0.1609) .. controls (-0.4484,-0.1875) and (-0.4672,-0.2266) .. (-0.4688,-0.2656) -- (-0.4688,-0.5312) -- (-0.4625,-0.5359) .. controls (-0.3891,-0.5844) and (-0.1922,-0.6094) .. (0.0047,-0.6094) .. controls (0.2016,-0.6094) and (0.3969,-0.5844) .. (0.4625,-0.5359) -- (0.4688,-0.5312) -- (0.4688,-0.2656) -- cycle
    (-0.1750,-0.0687) -- (-0.1484,-0.0578) .. controls (-0.1469,-0.0578) and (-0.1453,-0.0563) .. (-0.1438,-0.0563) .. controls (-0.1422,-0.0563) and (-0.1422,-0.0563) .. (-0.1406,-0.0547) -- (-0.1406,-0.0547) .. controls (-0.1203,-0.0437) and (-0.1094,-0.0234) .. (-0.1094,0.0000) -- (-0.1094,0.0281) .. controls (-0.0766,0.0109) and (-0.0391,0.0016) .. (0.0000,0.0016) .. controls (0.0391,0.0016) and (0.0766,0.0109) .. (0.1094,0.0281) -- (0.1094,0.0000) .. controls (0.1094,-0.0234) and (0.1219,-0.0437) .. (0.1422,-0.0547) -- (0.1422,-0.0547) .. controls (0.1422,-0.0547) and (0.1438,-0.0547) .. (0.1438,-0.0547) .. controls (0.1453,-0.0563) and (0.1469,-0.0563) .. (0.1484,-0.0578) -- (0.1750,-0.0687) -- (0.1719,-0.0813) .. controls (0.1375,-0.1078) and (0.0719,-0.1250) .. (0.0000,-0.1250) .. controls (-0.0719,-0.1250) and (-0.1391,-0.1078) .. (-0.1719,-0.0813) -- (-0.1750,-0.0687) -- cycle
    (0.0000,0.0312) .. controls (-0.1125,0.0312) and (-0.2031,0.1219) .. (-0.2031,0.2344) -- (-0.2031,0.2969) .. controls (-0.1406,0.3000) and (-0.0547,0.3234) .. (-0.0062,0.3438) -- (0.0094,0.3516) .. controls (0.0547,0.3750) and (0.0859,0.3891) .. (0.1266,0.4281) .. controls (0.1297,0.4094) and (0.1391,0.3484) .. (0.1531,0.3219) .. controls (0.1687,0.2953) and (0.1859,0.2828) .. (0.1969,0.2750) .. controls (0.1984,0.2734) and (0.2000,0.2734) .. (0.2016,0.2719) -- (0.2016,0.2344) .. controls (0.2031,0.1219) and (0.1125,0.0312) .. (0.0000,0.0312) -- cycle
    (0.2344,0.2812) .. controls (0.2344,0.2844) and (0.2328,0.2891) .. (0.2297,0.2922) .. controls (0.2266,0.2969) and (0.2203,0.3000) .. (0.2156,0.3031) .. controls (0.2047,0.3094) and (0.1922,0.3187) .. (0.1813,0.3391) .. controls (0.1719,0.3547) and (0.1625,0.3984) .. (0.1578,0.4344) .. controls (0.1562,0.4406) and (0.1531,0.4469) .. (0.1500,0.4516) .. controls (0.1438,0.4578) and (0.1359,0.4609) .. (0.1281,0.4609) .. controls (0.1281,0.4609) and (0.1281,0.4609) .. (0.1281,0.4609) .. controls (0.1203,0.4609) and (0.1125,0.4578) .. (0.1078,0.4531) .. controls (0.0687,0.4188) and (0.0422,0.4047) .. (-0.0031,0.3828) -- (-0.0188,0.3750) .. controls (-0.0672,0.3531) and (-0.1594,0.3297) .. (-0.2188,0.3297) .. controls (-0.2281,0.3297) and (-0.2344,0.3234) .. (-0.2344,0.3141) .. controls (-0.2344,0.3141) and (-0.2344,0.3141) .. (-0.2344,0.3141) -- (-0.2344,0.3141) -- (-0.2344,0.2359) .. controls (-0.2344,0.1594) and (-0.1969,0.0922) .. (-0.1406,0.0484) -- (-0.1406,0.0000) .. controls (-0.1406,-0.0125) and (-0.1484,-0.0234) .. (-0.1594,-0.0281) .. controls (-0.2016,-0.0469) and (-0.2641,-0.0250) .. (-0.2922,-0.0125) .. controls (-0.2781,0.0078) and (-0.2672,0.0312) .. (-0.2625,0.0594) .. controls (-0.2578,0.0906) and (-0.2609,0.1344) .. (-0.2641,0.1891) .. controls (-0.2672,0.2359) and (-0.2703,0.2875) .. (-0.2703,0.3484) .. controls (-0.2703,0.4594) and (-0.2469,0.5203) .. (-0.1906,0.5563) .. controls (-0.1234,0.5938) and (-0.0469,0.6047) .. (0.0281,0.5844) .. controls (0.1141,0.5578) and (0.1422,0.5094) .. (0.1422,0.5094) .. controls (0.1453,0.5031) and (0.1531,0.5000) .. (0.1594,0.5016) .. controls (0.1797,0.5062) and (0.2141,0.5094) .. (0.2266,0.5016) .. controls (0.2609,0.4828) and (0.2812,0.4562) .. (0.2812,0.3812) .. controls (0.2812,0.3312) and (0.2688,0.2672) .. (0.2578,0.2109) .. controls (0.2484,0.1609) and (0.2406,0.1188) .. (0.2422,0.0938) .. controls (0.2453,0.0500) and (0.2734,0.0109) .. (0.2906,-0.0094) .. controls (0.2625,-0.0234) and (0.2031,-0.0500) .. (0.1609,-0.0297) .. controls (0.1500,-0.0250) and (0.1438,-0.0141) .. (0.1438,-0.0016) -- (0.1438,0.0469) .. controls (0.2000,0.0891) and (0.2375,0.1578) .. (0.2375,0.2344) -- (0.2344,0.2812) -- cycle
    (-0.4375,-0.2656) .. controls (-0.4375,-0.2344) and (-0.4219,-0.2063) .. (-0.3969,-0.1828) .. controls (-0.3531,-0.1453) and (-0.2906,-0.1125) .. (-0.2047,-0.0797) -- (-0.0672,-0.5781) .. controls (-0.2250,-0.5734) and (-0.3734,-0.5516) .. (-0.4375,-0.5156) -- (-0.4375,-0.2656) -- cycle
    (-0.0344,-0.5781) -- (-0.1594,-0.1250) .. controls (-0.1188,-0.1453) and (-0.0609,-0.1562) .. (0.0000,-0.1562) .. controls (0.0609,-0.1562) and (0.1188,-0.1453) .. (0.1594,-0.1250) -- (0.0344,-0.5781) .. controls (0.0109,-0.5797) and (-0.0109,-0.5781) .. (-0.0344,-0.5781) -- cycle
    (0.0672,-0.5781) -- (0.2047,-0.0813) .. controls (0.2906,-0.1141) and (0.3531,-0.1469) .. (0.3969,-0.1844) .. controls (0.4219,-0.2063) and (0.4375,-0.2344) .. (0.4375,-0.2656) -- (0.4375,-0.5156) .. controls (0.3781,-0.5531) and (0.2281,-0.5734) .. (0.0672,-0.5781) -- cycle;
  }
}
\tikzset{
  hrc icon device/.pic={
  \path[fill=hrcMagenta]
    (0.4531,-0.2187) -- (0.4531,0.3438) -- (-0.4531,0.3438) -- (-0.4531,-0.2189) -- cycle
    (-0.4219,0.3125) -- (0.4219,0.3125) -- (0.4219,-0.1874) -- (-0.4219,-0.1877) -- cycle
    (-0.5000,0.3594) .. controls (-0.5000,0.3766) and (-0.4860,0.3906) .. (-0.4688,0.3906) -- (0.4688,0.3906) .. controls (0.4860,0.3906) and (0.5000,0.3766) .. (0.5000,0.3594) -- (0.5000,-0.2815) -- (0.5312,-0.2815) -- (0.5312,0.3594) .. controls (0.5312,0.3939) and (0.5033,0.4219) .. (0.4688,0.4219) -- (-0.4688,0.4219) .. controls (-0.5033,0.4219) and (-0.5312,0.3939) .. (-0.5312,0.3594) -- (-0.5312,-0.2815) -- (-0.5000,-0.2815) -- cycle
    (0.0781,-0.3280) -- (0.0781,-0.3593) -- (-0.0781,-0.3593) -- (-0.0781,-0.3280) -- (-0.7188,-0.3280) -- (-0.7188,-0.3593) .. controls (-0.7187,-0.4024) and (-0.6838,-0.4374) .. (-0.6406,-0.4374) -- (0.6406,-0.4374) .. controls (0.6838,-0.4374) and (0.7187,-0.4024) .. (0.7188,-0.3593) -- (0.7188,-0.3280) -- cycle
    (0.6406,-0.4062) -- (-0.6406,-0.4062) .. controls (-0.6665,-0.4062) and (-0.6875,-0.3852) .. (-0.6875,-0.3593) -- (-0.1094,-0.3593) .. controls (-0.1099,-0.3760) and (-0.0969,-0.3900) .. (-0.0802,-0.3905) .. controls (-0.0795,-0.3906) and (-0.0788,-0.3906) .. (-0.0781,-0.3905) -- (0.0781,-0.3905) .. controls (0.0948,-0.3911) and (0.1088,-0.3780) .. (0.1094,-0.3613) .. controls (0.1094,-0.3607) and (0.1094,-0.3600) .. (0.1094,-0.3593) -- (0.6875,-0.3593) .. controls (0.6875,-0.3852) and (0.6665,-0.4062) .. (0.6406,-0.4062) -- cycle;
  }
}
\tikzset{
  hrc icon chatbot/.pic={
  \path[fill=hrcTeal]
    (0.3905,-0.1813) -- (0.3905,-0.0389) .. controls (0.3905,-0.0088) and (0.3661,0.0156) .. (0.3360,0.0156) -- (0.2655,0.0156) -- (0.2655,0.0938) -- (0.0884,0.0938) -- (0.0884,0.1406) -- (0.1345,0.1406) .. controls (0.1809,0.1407) and (0.2186,0.1783) .. (0.2187,0.2248) -- (0.2187,0.4002) .. controls (0.2186,0.4467) and (0.1809,0.4843) .. (0.1345,0.4844) -- (0.0155,0.4844) -- (0.0155,0.5335) .. controls (0.0490,0.5421) and (0.0691,0.5762) .. (0.0604,0.6096) .. controls (0.0518,0.6430) and (0.0177,0.6631) .. (-0.0157,0.6545) .. controls (-0.0491,0.6459) and (-0.0692,0.6118) .. (-0.0606,0.5784) .. controls (-0.0549,0.5563) and (-0.0377,0.5392) .. (-0.0157,0.5335) -- (-0.0157,0.4844) -- (-0.1347,0.4844) .. controls (-0.1811,0.4843) and (-0.2188,0.4467) .. (-0.2188,0.4002) -- (-0.2188,0.2248) .. controls (-0.2188,0.1783) and (-0.1811,0.1407) .. (-0.1347,0.1406) -- (-0.0834,0.1406) -- (-0.0834,0.0938) -- (-0.2657,0.0938) -- (-0.2657,0.0001) -- (-0.3649,0.0001) .. controls (-0.3777,0.0001) and (-0.3881,0.0105) .. (-0.3881,0.0233) -- (-0.3881,0.1657) .. controls (-0.3415,0.1744) and (-0.3106,0.2192) .. (-0.3193,0.2659) .. controls (-0.3222,0.2813) and (-0.3291,0.2956) .. (-0.3395,0.3073) -- (-0.3629,0.2865) .. controls (-0.3428,0.2639) and (-0.3449,0.2293) .. (-0.3674,0.2093) .. controls (-0.3900,0.1892) and (-0.4246,0.1913) .. (-0.4446,0.2139) .. controls (-0.4629,0.2345) and (-0.4631,0.2654) .. (-0.4449,0.2862) -- (-0.4685,0.3067) .. controls (-0.4997,0.2710) and (-0.4960,0.2167) .. (-0.4602,0.1855) .. controls (-0.4486,0.1754) and (-0.4345,0.1685) .. (-0.4194,0.1657) -- (-0.4194,0.0233) .. controls (-0.4193,-0.0067) and (-0.3950,-0.0311) .. (-0.3649,-0.0311) -- (-0.2657,-0.0311) -- (-0.2657,-0.3598) .. controls (-0.3389,-0.3641) and (-0.3947,-0.4270) .. (-0.3903,-0.5002) .. controls (-0.3862,-0.5702) and (-0.3281,-0.6250) .. (-0.2579,-0.6250) -- (0.2577,-0.6250) .. controls (0.3310,-0.6250) and (0.3904,-0.5655) .. (0.3904,-0.4922) .. controls (0.3903,-0.4220) and (0.3356,-0.3639) .. (0.2655,-0.3598) -- (0.2655,-0.0156) -- (0.3360,-0.0156) .. controls (0.3489,-0.0156) and (0.3593,-0.0260) .. (0.3593,-0.0389) -- (0.3593,-0.1813) .. controls (0.3126,-0.1899) and (0.2818,-0.2348) .. (0.2905,-0.2814) .. controls (0.2933,-0.2968) and (0.3003,-0.3111) .. (0.3107,-0.3228) -- (0.3340,-0.3020) .. controls (0.3140,-0.2794) and (0.3160,-0.2448) .. (0.3386,-0.2248) .. controls (0.3612,-0.2047) and (0.3958,-0.2068) .. (0.4158,-0.2294) .. controls (0.4341,-0.2500) and (0.4342,-0.2809) .. (0.4161,-0.3017) -- (0.4396,-0.3222) .. controls (0.4708,-0.2865) and (0.4671,-0.2322) .. (0.4313,-0.2010) .. controls (0.4197,-0.1909) and (0.4057,-0.1841) .. (0.3905,-0.1813) -- cycle
    (-0.0313,0.5938) .. controls (-0.0313,0.6110) and (-0.0174,0.6250) .. (-0.0001,0.6250) .. controls (0.0172,0.6250) and (0.0312,0.6110) .. (0.0312,0.5938) .. controls (0.0312,0.5765) and (0.0172,0.5625) .. (-0.0001,0.5625) .. controls (-0.0174,0.5625) and (-0.0313,0.5765) .. (-0.0314,0.5937) .. controls (-0.0314,0.5937) and (-0.0314,0.5937) .. (-0.0314,0.5938) -- cycle
    (-0.2345,-0.1730) -- (-0.1067,-0.1730) -- (-0.0558,-0.0648) -- (-0.0080,-0.2480) -- (0.0565,-0.1152) -- (0.0972,-0.2116) -- (0.1287,-0.1794) .. controls (0.1328,-0.1756) and (0.1382,-0.1733) .. (0.1439,-0.1730) -- (0.2342,-0.1730) -- (0.2342,-0.3594) -- (-0.2345,-0.3594) -- cycle
    (-0.1347,0.1719) .. controls (-0.1639,0.1719) and (-0.1876,0.1956) .. (-0.1876,0.2248) -- (-0.1876,0.4002) .. controls (-0.1876,0.4294) and (-0.1639,0.4531) .. (-0.1347,0.4531) -- (0.1345,0.4531) .. controls (0.1637,0.4531) and (0.1874,0.4294) .. (0.1874,0.4002) -- (0.1874,0.2248) .. controls (0.1874,0.1956) and (0.1637,0.1719) .. (0.1345,0.1719) -- (-0.1347,0.1719) -- cycle
    (-0.0522,0.1406) -- (0.0572,0.1406) -- (0.0572,0.0938) -- (-0.0522,0.0938) -- cycle
    (-0.0834,0.0625) -- (0.2343,0.0625) -- (0.2343,-0.1417) -- (0.1439,-0.1417) .. controls (0.1305,-0.1420) and (0.1177,-0.1471) .. (0.1078,-0.1562) -- (0.0585,-0.0395) -- (0.0009,-0.1583) -- (-0.0474,0.0266) -- (-0.1266,-0.1417) -- (-0.2345,-0.1417) -- (-0.2345,0.0625) -- cycle
    (0.3593,-0.4922) .. controls (0.3592,-0.5482) and (0.3138,-0.5937) .. (0.2577,-0.5938) -- (-0.2579,-0.5938) .. controls (-0.3140,-0.5938) and (-0.3595,-0.5483) .. (-0.3595,-0.4922) .. controls (-0.3595,-0.4361) and (-0.3140,-0.3906) .. (-0.2579,-0.3906) -- (0.2577,-0.3906) .. controls (0.3138,-0.3907) and (0.3592,-0.4361) .. (0.3593,-0.4922) -- cycle
    (0.1452,0.3125) .. controls (0.1452,0.3496) and (0.1151,0.3797) .. (0.0780,0.3797) .. controls (0.0409,0.3797) and (0.0108,0.3496) .. (0.0108,0.3125) .. controls (0.0108,0.2754) and (0.0409,0.2453) .. (0.0780,0.2453) .. controls (0.1151,0.2453) and (0.1452,0.2754) .. (0.1452,0.3125) -- cycle
    (0.0780,0.2656) .. controls (0.0521,0.2656) and (0.0311,0.2866) .. (0.0311,0.3125) .. controls (0.0311,0.3384) and (0.0521,0.3594) .. (0.0780,0.3594) .. controls (0.1039,0.3594) and (0.1249,0.3384) .. (0.1249,0.3125) .. controls (0.1249,0.2866) and (0.1039,0.2656) .. (0.0780,0.2656) -- cycle
    (-0.0782,0.3797) .. controls (-0.1153,0.3797) and (-0.1454,0.3496) .. (-0.1454,0.3125) .. controls (-0.1454,0.2754) and (-0.1153,0.2453) .. (-0.0782,0.2453) .. controls (-0.0411,0.2453) and (-0.0110,0.2754) .. (-0.0110,0.3125) .. controls (-0.0110,0.3496) and (-0.0411,0.3797) .. (-0.0782,0.3797) -- cycle
    (-0.0782,0.2656) .. controls (-0.1041,0.2656) and (-0.1251,0.2866) .. (-0.1251,0.3125) .. controls (-0.1251,0.3384) and (-0.1041,0.3594) .. (-0.0782,0.3594) .. controls (-0.0523,0.3594) and (-0.0314,0.3384) .. (-0.0314,0.3125) .. controls (-0.0314,0.2866) and (-0.0523,0.2656) .. (-0.0782,0.2656) -- cycle;
  }
}
\tikzset{
  hrc icon hrsystem/.pic={
  \path[fill=hrcGreen]
    (0.4531,-0.4844) -- (0.4531,0.4531) .. controls (0.4531,0.5445) and (0.2197,0.5938) .. (0.0000,0.5938) .. controls (-0.2197,0.5938) and (-0.4531,0.5445) .. (-0.4531,0.4531) -- (-0.4531,-0.4844) .. controls (-0.4531,-0.5757) and (-0.2197,-0.6250) .. (0.0000,-0.6250) .. controls (0.2197,-0.6250) and (0.4531,-0.5757) .. (0.4531,-0.4844) -- cycle
    (0.0000,0.5625) .. controls (0.2415,0.5625) and (0.4219,0.5047) .. (0.4219,0.4531) .. controls (0.4219,0.4016) and (0.2415,0.3438) .. (0.0000,0.3438) .. controls (-0.2415,0.3438) and (-0.4219,0.4016) .. (-0.4219,0.4531) .. controls (-0.4219,0.5047) and (-0.2415,0.5625) .. (0.0000,0.5625) -- cycle
    (-0.4219,0.3991) .. controls (-0.3511,0.3425) and (-0.1714,0.3125) .. (0.0000,0.3125) .. controls (0.1714,0.3125) and (0.3511,0.3425) .. (0.4219,0.3991) -- (0.4219,0.1406) .. controls (0.4219,0.0891) and (0.2415,0.0312) .. (0.0000,0.0312) .. controls (-0.2415,0.0312) and (-0.4219,0.0891) .. (-0.4219,0.1406) -- cycle
    (-0.4219,0.0866) .. controls (-0.3511,0.0300) and (-0.1714,0.0000) .. (0.0000,0.0000) .. controls (0.1714,0.0000) and (0.3511,0.0300) .. (0.4219,0.0866) -- (0.4219,-0.1719) .. controls (0.4219,-0.2234) and (0.2415,-0.2812) .. (0.0000,-0.2812) .. controls (-0.2415,-0.2812) and (-0.4219,-0.2234) .. (-0.4219,-0.1719) -- cycle
    (-0.4219,-0.4844) -- (-0.4219,-0.2259) .. controls (-0.3511,-0.2825) and (-0.1714,-0.3125) .. (0.0000,-0.3125) .. controls (0.1714,-0.3125) and (0.3511,-0.2825) .. (0.4219,-0.2259) -- (0.4219,-0.4844) .. controls (0.4219,-0.5359) and (0.2415,-0.5938) .. (0.0000,-0.5938) .. controls (-0.2415,-0.5938) and (-0.4219,-0.5359) .. (-0.4219,-0.4844) -- cycle;
  }
}
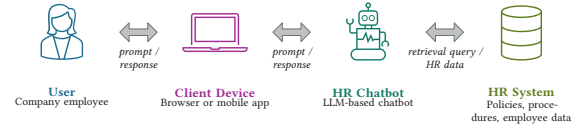
\begin{figure}
  \centering
\resizebox{0.95\linewidth}{!}{\begin{tikzpicture}[
  x=1cm, y=1cm,
  every node/.style={inner sep=0pt, outer sep=0pt},
  hrc title/.style={font=\small\bfseries, text width=3.3cm, align=center, anchor=north},
  hrc sub/.style={font=\footnotesize, text width=3.3cm, align=center, anchor=north,
                  text=hrcText, yshift=-1pt},
  hrc flow/.style={font=\footnotesize\itshape, text width=2.7cm, align=center, anchor=north,
                   text=hrcText},
]

\pic at (0.000,0) {hrc icon user};
\pic at (3.375,0) {hrc icon device};
\pic at (6.750,0) {hrc icon chatbot};
\pic at (10.125,0) {hrc icon hrsystem};

\foreach \ax in {1.6875, 5.0625, 8.4375} {
  \path[fill=hrcArrow, draw=hrcArrowEdge, line width=0.4pt, line join=round]
    (\ax-0.4125,0) -- (\ax-0.2063,0.2063) -- (\ax-0.2063,0.0722)
    -- (\ax+0.2063,0.0722) -- (\ax+0.2063,0.2063) -- (\ax+0.4125,0)
    -- (\ax+0.2063,-0.2063) -- (\ax+0.2063,-0.0722)
    -- (\ax-0.2063,-0.0722) -- (\ax-0.2063,-0.2063) -- cycle;
}

\node[hrc flow] at (1.6875,-0.34) {prompt /\\response};
\node[hrc flow] at (5.0625,-0.34) {prompt /\\response};
\node[hrc flow] at (8.4375,-0.34) {retrieval query /\\HR data};

\node[hrc title, text=hrcBlue]    (tu) at (0.000,-1.22)  {User};
\node[hrc title, text=hrcMagenta] (td) at (3.375,-1.22)  {Client Device};
\node[hrc title, text=hrcTeal]    (tc) at (6.750,-1.22)  {HR Chatbot};
\node[hrc title, text=hrcGreen]   (th) at (10.125,-1.22) {HR System};

\node[hrc sub] at (tu.south) {Company employee};
\node[hrc sub] at (td.south) {Browser or mobile app};
\node[hrc sub] at (tc.south) {LLM-based chatbot};
\node[hrc sub] at (th.south) {Policies, procedures, employee data};

\end{tikzpicture}}
\caption{Running example HR chatbot system}\label{fig:running-example}
\end{figure} 
In this example, employees interact with the chatbot to obtain information about HR matters (e.g., holiday policies), while the underlying LLM generates responses from the information available to it. 
Such an LLM-based system introduces privacy threats that do not fit neatly into mitigation guidance developed for systems that primarily collect, store, and process predefined personal data, mainly because of the opaque, non-deterministic, and generative nature of LLMs.

To illustrate, we focus on two related LINDDUN4GenAI~\cite{liao2026} threat characteristics. 
The first is \emph{DD.3.5 Fabrication}, in which the GenAI system generates inaccurate or fabricated personal data. 
In the HR-chatbot scenario, the model may therefore produce a statement about an employee that is not grounded in an actual record. 
Unlike conventional data disclosure, the problematic personal data need not have existed in the system before the response was generated.

The second is \emph{U.2.3.1 Inability to rectify/erase hallucinated personal data}. 
Once such personal data have been hallucinated, the affected data subject may be unable to rectify or erase them because there is no corresponding stored record to modify or delete. 
The problem is therefore not simply that incorrect personal data exist, but that conventional mechanisms for intervening in stored data may not apply to content generated dynamically by an LLM.

\subsection{Sub-problems}
\label{subsec:subproblems}
We decompose the resulting gap into three sub-problems.
The HR-chatbot example introduced above is used throughout to illustrate how each limitation manifests in practice.

\subsubsection*{Sub-problem 1: Lack of fine-grained threat-to-mitigation mapping for GenAI system.}
Existing threat-to-mitigation guidance operates at the level of top-level threat types or high-level design strategies~\cite{hoepman2014,hoepman2022,colesky2016}, discarding the fine-grained knowledge encoded in the lower nodes of the threat trees. 
Al-Momani et al.~\cite{almomani2022} 
provide finer-grained threat-to-mitigation guidance for classical LINDDUN by identifying \emph{key nodes}: selected threat-tree nodes that preserve information about the cause of a threat and connect it to a corresponding mitigation goal. Their resulting flowcharts can therefore distinguish between different causes of threats belonging to the same high-level category. 
However, these key nodes and flowcharts were constructed before the GenAI-specific threat characteristics of LINDDUN4GenAI~\cite{liao2026} were introduced. 
No key nodes or solution paths therefore exist for these GenAI-specific threats.

Consider \emph{DD.3.5 Fabrication} in the HR chatbot. 
At the top-level data-disclosure category, the threat may appear similar to other situations involving personal data exposed by a system. 
At the lower level, however, its cause is fundamentally different: the problematic personal data are \emph{generated by the model} rather than retrieved from an existing data store or exposed through an insufficiently protected data flow. 
This distinction is crucial for mitigation selection. 
Guidance intended to protect existing personal data cannot simply be assumed to address personal data that the model invents at generation time.

The second running threat makes the limitation even clearer. 
\emph{U.2.3.1 Inability to rectify/erase hallucinated personal data} belongs to the soft privacy threat type of unawareness, which is outside the scope of the flowcharts of Al-Momani et al.~\cite{almomani2022}. 
Thus, while LINDDUN4GenAI can identify the precise unawareness problem, there is no existing bridge from threat to mitigation.

To recap, GenAI introduces novel threat characteristics that current mitigation-selection approaches (e.g., flowcharts designed by Al-Momani et al.~\cite{almomani2022}) do not address. 
Nor can these existing flowcharts be directly transferred to these new GenAI threats: their applicability questions (``are all used attributes required for the system?'', ``could sensitive attributes be replaced with different, yet less sensitive, attributes?''~\cite{almomani2022}) assume that the relevant attributes were deliberately collected by the engineer and are therefore ill-suited to generated content.

\subsubsection*{Sub-problem 2: Solution-space assumptions are inapplicable in the GenAI context.}
Even where potentially relevant mitigations exist, the assumptions underlying the existing solution-space do not always hold for GenAI systems.

First, the coverage asymmetry outlined above works against GenAI: the threat literature synthesized in LINDDUN4GenAI is dominated by leakage threats (memorization, extraction, and inference of training, fine-tuning, and prompt data) that map to hard privacy threat types~\cite{liao2026}, where the pattern catalog is lacking~\cite{almomani2021}. 

Second, the soft privacy patterns that do exist rest on an assumption that no longer holds in the GenAI context: that the system holds stored records that a data subject can access, rectify, or erase. 

The HR-chatbot example shows why this assumption fails. 
Under \emph{DD.3.5 Fabrication}, the chatbot may generate personal data that were never stored as a record. 
The affected employee subsequently encounters \emph{U.2.3.1 Inability to rectify/erase hallucinated personal data}, as for hallucinated personal data ``there are simply no records to `rectify' or `delete'\,'', as 
(i)~because the system is inherently probabilistic, removing one observed output also does not ensure that a similar statement will not be generated again, 
and (ii)~personal data absorbed into model weights cannot, in general, be selectively untrained (U.2.3.2)~\cite{liao2026}. 

Consequently, applying a conventional intervention mechanism without considering the nature of generated content may satisfy the intended pattern only superficially. 
For example, providing an interface through which an employee can inspect or request correction of stored HR data does not, by itself, explain how the organization should rectify or correct personal data that the chatbot dynamically fabricates.

Third, the pattern properties cataloged for selection support~\cite{almomani2021} omit properties that are particularly relevant to determining applicability in GenAI systems, most notably whether the deploying party controls the training data, the model weights, or only the prompts.
A GenAI threat-to-mitigation mapping must therefore do more than assign existing mitigations to new threats: it must specify the conditions under which a mitigation is applicable and make explicit where no adequate mitigation currently exists.  
Such GenAI-relevant properties can be found in more recent research, such as the work of Kunz and Binder~\cite{kunz2022} or Wang et al.~\cite{wang2025}.

\subsubsection*{Sub-problem 3: Prioritization difficulty under GenAI constraints.}
Even after applicable mitigations have been identified, engineers may still face several alternative mitigation strategies and need guidance on which one to consider first.

Currently, the mitigation strategies themselves come with no such ordering~\cite{hoepman2014,hoepman2022}. The only ordering available in the fine-grained bridge is the one Al-Momani et al.~\cite{almomani2022} adopt in their flowcharts: 
countermeasures of the \emph{minimize} strategy first, then \emph{separate}, \emph{abstract}, and \emph{hide}, so that options with little utility impact are examined before options carrying significant computational or communication overhead.
Al-Momani et al.\ nevertheless emphasize that this is one possible ordering rather than an intrinsic ordering of the strategies themselves~\cite{almomani2022}.
Furthermore, their ordering presumes that the threat concerns \emph{collected} data flowing through the system: each strategy reduces the amount, linkability, or precision of data already in flow.
It offers no guidance for threats concerning \emph{generated} content or \emph{influence on the data subject}.
For example, minimization asks whether particular attributes need to be collected, stored, or processed in the first place. This reasoning is difficult to apply directly to \emph{DD.3.5 Fabrication}. 
In the HR-chatbot example, the problematic personal data do not necessarily exist before generation: the model may produce them dynamically in response to an employee's query. 
There is therefore no predefined personal-data attribute that can simply be removed before it enters the system.

The same problem extends to \emph{U.3 Interference with personal decision making}, a soft privacy threat, which concerns the model's effect on the subject's decision-making rather than any data asset. 
In \emph{U.2.3.1 Inability to rectify/erase hallucinated personal data}, once the chatbot has hallucinated personal data, candidate interventions may act at very different points in the GenAI lifecycle or architecture. 
Their feasibility depends on factors such as whether the organization controls the underlying model, can modify its training or fine-tuning procedure, can only modify prompts and retrieved context, or consumes the model through a third-party API. 
An ordering designed around collected attributes does not capture these distinctions.
Neither of these unawareness threats is covered by the existing ordering, as Al-Momani et al.~\cite{almomani2022} explicitly scope out the soft privacy threat type in which U.2.3.1 and U.3 reside. 

Moreover, the deployment paradigm is also an additional prioritization constraint (pre-trained, fine-tuned, prompt-based, or agentic~\cite{liao2026}) since it determines which mitigations are feasible at all. 
A technique such as Differentially Private Stochastic Gradient Descent (DP-SGD)~\cite{abadi2016}, for instance, cannot be applied to a model consumed through a third-party API, since the consuming party using the model has no control over model training.
Conversely, mitigations operating at inference time may remain feasible in such a deployment.
While existing work such as that of Wang et al.~\cite{wang2025} organizes their GenAI risks and countermeasures based on this deployment paradigm, they do not use it to filter infeasible mitigations or to order the remaining ones.
 \section{Toward GenAI-Aware Mitigation Selection}\label{sec:approach}
The sub-problems of \cref{subsec:subproblems} delineate what any future mitigation-selection approach for GenAI privacy threats must accomplish. In this section, we first formulate the recommendations that candidate approaches should follow (\cref{sec:rec}), and then sketch one concrete direction that we consider promising (\cref{sec:cansol}), to illustrate that these recommendations are attainable.

\subsection{Recommendations for future approaches}\label{sec:rec}
We derive recommendations from the sub-problems identified in \cref{sec:problem}: R1 addresses sub-problem~1, R2 addresses sub-problem~2, and R3 addresses sub-problem~3, while R4 is a cross-cutting adoption concern. These recommendations are deliberately framework-agnostic: although this paper uses LINDDUN4GenAI as its anchor, they should hold for any approach that bridges a GenAI privacy threat knowledge base and a body of mitigations.

\begin{description}
  \item[R1: Fine-grained threat-to-mitigation guidance.] 
  Selection guidance must operate at the granularity at which threats are identified: at the level of threat causes rather than broad threat types. 
  Whatever the underlying threat knowledge base, be it the LINDDUN4GenAI threat trees~\cite{liao2026} or other GenAI threat taxonomies~\cite{shanmugarasa2025,ma2025,wang2025}, the fine-grained knowledge it captures must be used for mitigation selection rather than dismissed.
  Relying on key nodes~\cite{almomani2022} is one existing instantiation of this principle (sub-problem~1).

  \item[R2: Explicit applicability properties.] Candidate mitigations must be annotated with the properties that actually determine their applicability in GenAI systems, e.g., 
    (i)~whether the deploying party controls the training data, the model weights, or only the prompts; 
  (ii)~the impact on model utility; and 
  (iii)~the assumptions the mitigation relies on, such as the existence of identifiable, rectifiable records. 
  Where no adequate mitigation exists at all, this absence must be made explicit (sub-problem~2).

  \item[R3: GenAI-valid prioritization principles.] Ordering principles must hold in the GenAI context: 
  (i)~feasibility filtering by deployment paradigm before any ranking; 
  (ii)~a cost model reflecting GenAI-specific overhead; and 
  (iii)~dedicated treatment of threats concerning generated content and influence on the data subject, for which minimization-first logic does not apply (sub-problem~3).
\item[R4: Compatibility and evolvability.] To be adopted, the guidance should integrate with established privacy engineering and threat modeling practice rather than require a parallel process, and it should remain maintainable as models, deployment patterns, and attacks evolve, for instance through an extensible, publicly maintained knowledge base as exemplified by LINDDUN4GenAI~\cite{liao2026}.
\end{description}

\subsection{A candidate direction: extending the key-node method}\label{sec:cansol}
The construction approach of Al-Momani et al.~\cite{almomani2022} is generic by design: it prescribes how to build selection support from a threat knowledge base.
One promising direction is therefore to re-instantiate this approach on the LINDDUN4GenAI threat trees, with GenAI-specific modifications at each step.

First, key nodes would be identified for the GenAI-specific threat characteristics (R1). 
The lower nodes of the new trees contain the cause-related information required for the identification of key-nodes.
The children of \emph{DD.1.3 Data type structure}, for instance, distinguish leakage through the data encoding itself from leakage through internal derivations such as embeddings and gradients~\cite{liao2026}.

Second, the resulting key nodes would be grouped into mitigation goals. 
We anticipate that some goals will have no classical counterpart, such as preventing or containing fabricated personal data (DD.3.5) and constraining the system's influence on the data subject (U.3), and that grouping may need to span CAMs rather than DFD elements alone.

Third, candidate mitigations would be assembled for each goal from the existing solution-space layers (strategies and tactics~\cite{hoepman2014,hoepman2022,colesky2016}, patterns~\cite{privacypatterns}, PETs~\cite{heurix2015}) and from GenAI-specific techniques cataloged in recent surveys~\cite{shanmugarasa2025}, each annotated with the applicability properties and explicit gap flags where no adequate option exists (R2).

Fourth, the ordering step would be revised (R3). 
For hard privacy threats concerning collected data, the utility-loss rationale behind the existing minimize-first ordering may be retained, though recalibrated to GenAI cost structures. Homomorphic encryption, for example, often comes at a considerable performance decrease, which may not be suitable for LLM inference as computational power demands and processing times would be prohibitively large for the use case~\cite{shah2026}. 
Before any ranking, a feasibility filter based on the deployment paradigm could discard inapplicable mitigations. 
For generated content and influence threats (DD.3.5, U.3), dedicated orderings would be needed, grounded in the deployment paradigm and in the nature of the threat actor, whether a user, the system itself, or an agent.

Finally, applicability questions and output representations analogous to the existing solution flowcharts~\cite{almomani2022} would keep the result recognizable to analysts already familiar with that method (R4).

We emphasize that this is an open-ended direction: constructing the mapping, defining the heuristics, and validating both with practitioners remain open research tasks for future work.
 \section{Conclusion}\label{sec:conclusion}
Privacy threat modeling for GenAI-based systems and the development of GenAI privacy mitigations are both advancing, but as separate, still-maturing bodies of knowledge: matching the right mitigation to an identified threat remains unsupported. 

LINDDUN4GenAI~\cite{liao2026} provides fine-grained, GenAI-specific problem-space knowledge, yet 
(i)~there is no fine-grained guidance that connects GenAI-specific threat knowledge to fitting mitigations;
(ii)~traditional solution-space assumptions do not hold up
in the GenAI context; and 
(iii)~there is no ordering principle valid under GenAI constraints.

We argue that closing this gap is a research task for the privacy engineering community, not an exercise in applying existing methods. 
To that end, we formulated four recommendations that future mitigation-selection approaches should satisfy and sketched one candidate direction: extending the key-node method of Al-Momani et al.~\cite{almomani2022} to the LINDDUN4GenAI threat trees, with mitigation goals, applicability properties, and orderings revised for GenAI.\ 
Elaborating this direction, constructing the actual threat-to-mitigation mapping, and validating it in practice fall outside the scope of a position paper and are left as future work.

Closing the gap will also require expertise beyond computer science. 
Data protection principles such as storage limitation and minimization must be interpreted in ways that are technically meaningful for models that memorize their training data, and regulatory expectations under frameworks such as the EU AI Act~\cite{EU2024ai} must be connectable to concrete mitigations.
Without such interdisciplinary work, data subject rights risk remaining formal: acknowledged in policy, but not enforced at the level of system design. By articulating this absence of mitigation-selection guidance for GenAI privacy threats, we aim to motivate research that bridges the gap.

\appendix

\begin{acks}
This research is partially funded by the Research Fund KU Leuven, Internal Funds KU Leuven, and by the Cybersecurity Research Program Flanders.
\end{acks}

\begin{ai}
GenAI-based tools were used to revise the text, improve flow, and correct any typos, grammatical errors, and awkward phrasing. 
\end{ai}

\bibliographystyle{ACM-Reference-Format}
\bibliography{references}

\end{document}